\documentclass[journal,twoside,web]{ieeecolor}
\usepackage{lcsys}
\usepackage{amsmath,amssymb,amsfonts}
\usepackage{algorithmic}
\usepackage{graphicx}
\usepackage{textcomp}
\def\BibTeX{{\rm B\kern-.05em{\sc i\kern-.025em b}\kern-.08em
    T\kern-.1667em\lower.7ex\hbox{E}\kern-.125emX}}
    
\usepackage{epsfig} 

\usepackage{amsthm}

\usepackage{bm}

\usepackage{accents}
\usepackage[noadjust]{cite}
\usepackage{hyperref}
\usepackage[outdir=Figures/]{epstopdf}
\usepackage{wrapfig}
\usepackage{float}
\usepackage{capt-of}
\usepackage{cuted}

\newcommand{\R}{\mathbb{R}}

\newcommand{\Xcal}{\mathcal{X}}

\newcommand{\Scal}{\mathcal{S}}

\newcommand{\Dcal}{\mathcal{D}}
\newcommand{\Ccal}{\mathcal{C}}
\newcommand{\Hcal}{\mathcal{H}}

\newcommand{\Rplus}{\R_{\geq 0}}

\newcommand{\classKinfty}{\mathcal{K}_{\infty}}
\newcommand{\classKL}{\mathcal{KL}}

\definecolor{darkblue}{RGB}{0,0,102}
\definecolor{lightblue}{RGB}{77,77,148}

\definecolor{gold}{RGB}{234, 170, 0}
\definecolor{metallic_gold}{RGB}{139, 111, 78}

\newcommand{\norm}[1]{\left\Vert #1 \right\Vert}

\newcommand{\lmat}{\begin{bmatrix}}
\newcommand{\rmat}{\end{bmatrix}}

\newcommand{\eqn}[1]{\begin{align}#1\end{align}}
\newcommand{\eqnN}[1]{\begin{align*}#1\end{align*}}

\newcommand{\half}{\frac{1}{2}}
\newcommand{\bmat}[1]{\begin{bmatrix}#1\end{bmatrix}}

\newtheorem{theorem}{Theorem}

\newtheorem{lemma}{Lemma}
\newtheorem{assumption}{Assumption}

\theoremstyle{definition}
\newtheorem{example}{Example}
\newtheorem{remark}{Remark}
\newtheorem{definition}{Definition}

\newif\ifshowImages
\showImagesfalse
\showImagestrue

\newif\iffast
\fastfalse
\fasttrue

\begin{document}
\title{
A Constructive Method for Designing Safe Multirate Controllers for Differentially-Flat Systems
}
\author{
Devansh R. Agrawal$^{\dagger, 1}$, Hardik Parwana$^{\dagger, 1}$, Ryan K. Cosner$^{\dagger, 2}$,  Ugo Rosolia$^{2}$, Aaron D. Ames$^2$, Dimitra Panagou$^1$%
\thanks{The authors would like to acknowledge the support of the Office of Naval Research (ONR), under grant number N00014-20-1-2395, and of the National Science Foundation (NSF) under grant no. \#1942907 and \#1931982. The views and conclusions contained herein are those of the authors only and should not be interpreted as representing those of ONR, the U.S. Navy or the U.S. Government. This work is also partially supported by BP and NSF Award \#1932091}%
\thanks{$\dagger$ These authors contributed equally to this work.}
\thanks{$^{1}$Authors are with University of Michigan, Ann Arbor, USA.
        {\tt\small \{devansh, hardiksp, dpanagou\}@umich.edu}}%
\thanks{$^{2}$Authors are with California Institute of Technology, California, USA.
        {\tt\small \{rkcosner, urosolia, ames\}@caltech.edu}}%
}

\maketitle
\thispagestyle{empty}


\ifshowImages
\begin{strip}
    \vspace{-120pt}
    \centering
    \includegraphics[width=\linewidth]{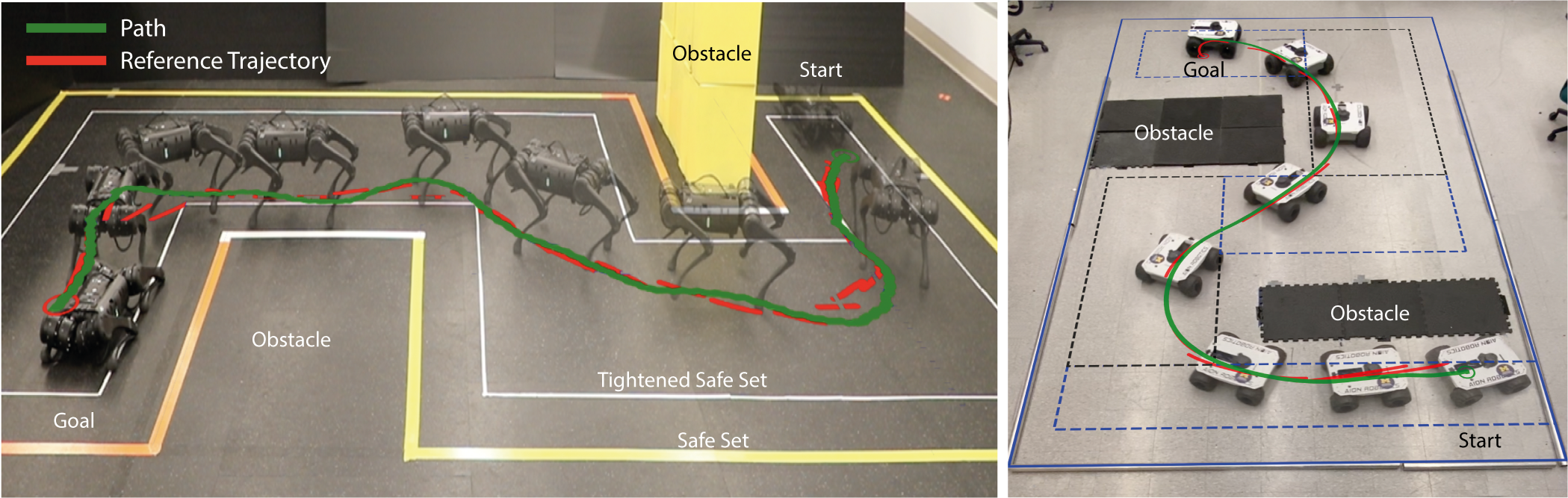}
    \vspace{-15pt}
    \captionof{figure}{ Snapshots of a quadrupedal robot (left) and ground rover (right) navigating safely from start to goal positions around two rectangular obstacles. The safe set (thick outside line) and the tightened set (thin/dashed lines) are shown. The reference trajectory (red) is solved online using Model Predictive Control, and must lie inside the tightened safe set. A tracking controller ensures the maximum deviation from this reference trajectory is smaller than the tightening. Thus the true path (green) remains within the safe set. Video: \url{https://tinyurl.com/3c58cnjj}.}
    \label{fig:experiment_scenario}
    \vspace{-5pt}
\end{strip}
\fi

\begin{abstract}
We present a multi-rate control architecture that leverages fundamental properties of differential flatness to synthesize controllers for safety-critical nonlinear dynamical systems. We propose a two-layer architecture, where the high-level generates reference trajectories using a linear Model Predictive Controller, and the low-level tracks this reference using a feedback controller. The novelty lies in how we couple these layers, to achieve formal guarantees on recursive feasibility of the MPC problem, and safety of the nonlinear system. Furthermore, using differential flatness, we provide a constructive means to synthesize the multi-rate controller, thereby removing the need to search for suitable Lyapunov or barrier functions, or to approximately linearize/discretize nonlinear dynamics. We show the synthesized controller is a convex optimization problem, making it amenable to real-time implementations. The method is demonstrated experimentally on a ground rover and a quadruped robotic system.
\end{abstract} 

\section{INTRODUCTION}

\IEEEPARstart{C}{ontrol} of nonlinear systems for navigating a constrained environment is a common problem in safety-critical robotics. Despite the extensive work on planning and control methods, the real-time deployment of controllers that provide guarantees of safety poses challenges either due to the computational complexity of nonlinear and nonconvex optimization, or due to the curse of dimensionality in search based approaches~\cite{bansal2017hamilton}.

Recently, multirate controllers have shown promising results for combining planning and tracking~\cite{gurriet2018towards, ames2019control, wang2017safety, herbert2017fastrack,yin2020optimization, singh2018robust, gao2014tube, kogel2015discrete, yu2013tube,kohler2020computationally, rosolia2020multi, garg2021multi}. Safety can be guaranteed using low-level filters that compute the closest safe control action to a desired command using control barrier functions~\cite{gurriet2018towards,ames2019control,wang2017safety}. The tracking error and control policy can be computed using Hamilton-Jacobi (HJ) reachability analysis~\cite{herbert2017fastrack} or sum-of-squares programming~\cite{singh2018robust,yin2020optimization}. Nonlinear tube MPC approaches have been developed to bridge high-level planning with low-level control~\cite{gao2014tube, kogel2015discrete, yu2013tube,kohler2020computationally, rosolia2020multi, garg2021multi}. However the nonlinearity in the MPC poses a challenge in identifying suitable barrier functions in the tracking layer. In~\cite{fan2020fast}, a tube-based planning approach with safety guarantees was developed.
However since the size of the tube increases over the planning horizon, the approach is conservative and cannot be used for in recursive planning methods like MPC.

Many dynamical systems, including unicycles, quadrotors, inverted pendulums, and induction motors, possess a useful property known as differential flatness (see \cite[Ch. 7]{martin2003flat} for a catalog of flat systems).  Several studies have identified that such systems possess useful properties for planning~\cite{wang2017safety, murray1995differential, fliess1995flatness, martin2003flat, mellinger2011minimum}. For instance in~\cite{mellinger2011minimum}, quadrotor trajectories from initial to target flat states are represented as polynomials whose coefficients are determined by solving a linear system. However, the method does not provide a principled way of incorporating disturbances or safety constraints.

In this work, we propose a novel multirate controller that leverages properties of differentially-flat systems. These properties allow a constructive means of designing both the planner, a linear MPC, and the tracker, an Input-to-State  Stable (ISS) feedback controller. We provide formal guarantees of safety for the continuous-time (low-level) system, and recursive feasibility for the MPC planner. We experimentally demonstrate our framework on a ground rover and a quadrupedal robot that can be modeled as unicycles. 

\section{PRELIMINARIES}
\label{section::preliminaries}

Let $\R$ denote the set of reals and $\Rplus$ the set of non-negative reals. The $p$-norm of a vector is $\norm{\cdot}_p$ and $p=2$ when the subscript is omitted. For a signal $d: \Rplus \to \R^n$, let the $\mathcal{L}_\infty$ norm be $\norm{d(t)}_\infty \triangleq \sup_{t\geq 0}\{\norm{d(t)}\}$. If the signal is bounded, i.e., $\norm{d(t)}_\infty < \infty$, it belongs to the set of bounded signals $\mathbb{L}^n_\infty$.
The min, max eigenvalues of a symmetric matrix $M \in \R^{n\times n}$ are $\lambda_{min}(M)$, $\lambda_{max}(M)$. 
The set of $\classKinfty$ and $\classKL$ functions are as defined in~\cite{ames2019control}. $L_fV(x)$ denotes the Lie derivative of $V:\R^n\to\R$ along $f:\R^n \to \R^n$ evaluated at $x\in\R^n$. For two sets $\mathcal{C}, \mathcal{D} \subset \mathbb{R}^n$, $\mathcal{C} \ominus \mathcal{D} = \{c : c + d \in \mathcal{C} \ \forall d \in \mathcal{D}\}$ denotes the Pontryagin set difference.

\subsection{Differentially-Flat Systems}

Consider a control-affine nonlinear system 
\eqn{
\dot x = f(x) + g(x)u
\label{eqn:control-affine-system}
}
where $x \in \Xcal \subset \R^n$ is the state, and $u \in \R^m$ is the control input. The functions $f: \Xcal \to \R^n$ and $g: \Xcal \to \R^{n \times m}$ are smooth and satisfy $f(0) = 0$. Differential flatness of such systems is often defined in terms of flat outputs (as in \cite{mellinger2011minimum}) or using differential geometry (e.g.~\cite{fliess1999lie, martin2003flat}). Following \cite{levine2007equivalence}, here we define it in terms of endogenous dynamic feedback.\footnote{This is possible since \cite[Thm.~3]{levine2007equivalence} shows that a system is differentially flat if and only if it admits an endogenous dynamic feedback. See \cite[Sec 5.3.6]{levine2009analysis} for an explanation of the term \emph{endogenous}.}
\begin{definition}
The control system \eqref{eqn:control-affine-system} is \emph{differentially flat} over a domain $\mathcal{M} \subset \Xcal \times \R^q$ if (I) there exists an \emph{endogenous dynamic feedback}
\eqn{
\dot y &= a(x,y,v) \label{eqn:dynfeedbackA}\\
u &= b(x,y,v), \label{eqn:dynfeedbackB}
}
where $y \in \R^q$ is an additional state (referred to as the \emph{dynamic extension}) and $v \in \R^m$ is a different control input (referred to as the \emph{flat control input}, and (II) there exists a diffeomorphism $\Xi: \mathcal{M} \to \mathcal{N}$, where $\mathcal{N} \subset \R^n \times \R^q$ is a domain, 
\eqn{
\xi = \Xi(x,y), \label{eqn:Xi}
}
which maps the nonlinear states $(x,y)$ to the \emph{flat state} $\xi \in \R^{n+q}$ such that the dynamics of $\xi$ are linear, time-invariant
\eqn{
\dot \xi = \frac{\partial \Xi}{\partial x} \dot x + \frac{\partial \Xi}{\partial y} \dot y 
= A \xi + B v, \label{eqn:flat_system}}
where $A, B$ are constant matrices of appropriate size.
\end{definition}
\begin{remark}
When a system is differentially flat, the flat system is linear, time-invariant, and controllable~\cite{fliess1999lie}.
\end{remark}
\begin{remark}
If $q = 0$ (the dimension of $y$), the dynamic feedback is equivalent to a full-state feedback, and thus full-state feedback linearizable systems are a subset of differentially-flat systems (subject to smoothness requirements) \cite{ramasamy2014dynamically}.
\end{remark}

\begin{example}[Unicycle: Differential flatness]
The unicycle is a differentially-flat system with nonlinear dynamics
\eqnN{
\dot x_1 = u_1 \cos x_3, \quad \dot x_2 = u_2 \sin x_3, \quad \dot x_3 = u_2,
}
where $(x_1, x_2)$ is the position, $x_3$ is the heading angle, and the control inputs are linear and angular velocities, $u_1, u_2$.
The dynamic extension is ${y = [\dot x_1, \dot x_2]^T}$, and flat state is ${\xi \in \R^4}$. Then ${\Xi(x,y) = [x_1, x_2, y_1, y_2]^T}$. 
The flat system dynamics are
\eqnN{\dot \xi_1 = \xi_3, \quad \dot \xi_2 = \xi_4, \quad \dot \xi_3 = v_1, \quad \dot \xi_4 = v_2,}
where $v \in \R^2$ is the flat control input. The nonlinear state and control input can be determined from $\xi$ and $v$ as
\eqnN{
x_1 = \xi_1, \quad x_2= \xi_2,\quad  x_3 = \arctan{(\xi_4 / \xi_3)},\\
u_1 = \sqrt{\xi_3^2 + \xi_4^2}, \quad u_2= \frac{-\xi_4 v_1 + \xi_3 v_2}{\xi_3^2 + \xi_4^2}.
}
Notice that $\Xi^{-1}$ has a singularity at $\dot x_1 = \dot x_2 = 0$, and thus is excluded from the domain $\mathcal{M}$. See Example 2 and \cite[Sec. 2.5]{martin2003flat} on methods to handle singularities.
\end{example}

\subsection{Input-To-State Stability}

Assume that a bounded additive disturbance\footnote{ISS is typically defined for \emph{matched disturbances} $w = g(x) d$ \cite{sontag2008input}.  In this paper, we consider the \emph{unmatched case} as necessitated by the coordinate transformations resulting from differential flatness.}  $w : \Rplus \to \R^n$, $w \in \mathbb{L}^n_\infty$ is introduced to the system \eqref{eqn:control-affine-system}, and that a feedback controller $\pi:\Xcal \to \R^m$, $\pi(0) = 0$ has been designed. Then the closed-loop system dynamics read: 
\eqn{
\dot x = f(x) + g(x) \pi(x) + w(t),
\label{eqn:disturbed_control_affine_sys}
}
where $\norm{w(t)}_\infty \triangleq \sup_{t\geq 0}\{|w(t)|\} \leq \bar w$ for some $\bar w < \infty$. 

\begin{definition}
A controller $\pi: \Xcal \to \R^m$, $\pi(0) = 0$ and system \eqref{eqn:disturbed_control_affine_sys} are \emph{input-to-state stabilising} and \emph{input-to-state stable}, respectively, wrt. $w$, if there $\exists$ $\beta \in \classKL$, $\iota \in \classKinfty$ s.t.
\eqn{
\norm{x(t, x_0, w)} \leq \beta(\norm{x_0}, t) + \iota(\norm{w}_\infty)
}
for all $x_0 \in \Xcal, w \in \mathbb{L}^n_\infty$ and $t\geq 0$.
\end{definition}

\begin{definition}
A continuously differentiable positive definite function $V: \R^n \to \Rplus$ is an \emph{input-to-state stabilising control Lyapunov function (ISS-CLF)} with respect to $w$, if there exists functions $\alpha, \iota \in \classKinfty$ such that $\forall x \in \Xcal$ and $w \in \mathbb{L}^n_\infty$, 
\eqn{
\inf_{u\in \mathbb{R}^m} \left[L_fV(x) + L_gV(x) u + \frac{\partial V}{\partial x}w \right] \leq -\alpha(\norm{x}) + \iota(\norm{w}_\infty).
\label{eqn:ISSCLF_def}
}
\end{definition}

\begin{definition}
For the system \eqref{eqn:disturbed_control_affine_sys}, a set $\Dcal \subset \R^n$ is \emph{robustly control invariant}, with respect to disturbances $w$ bounded by $\bar w$,  if there exists a feedback controller $\pi(x) \in \R^m$ such that $x(t_0) \in \Dcal \implies x(t) \in \Dcal$ for all $t\geq t_0$ and for all $w \in \mathbb{L}^m_\infty$ where $\norm{w(t)}_\infty \leq \bar w$.
\end{definition}

\section{CONTROLLER CONSTRUCTION}
\label{section::controllers}

Our multirate controller consists of two stages: A high-level planner, in the form of a linear MPC, and a low-level tracker, in the form a feedback controller. In designing the low-level tracker, we define explicitly a set $\Dcal$, which is the set of possible tracking errors between the reference and current state. In the high-level, we shrink the safe set by $\Dcal$ and require the MPC to generate reference trajectories that lie within the tightened safe set. As a consequence, the system's trajectory will lie in the safe set, despite the disturbances.

\subsection{High-Level Planning}

At the high-level, we solve a Finite Time Optimal Control Problem (FTOCP) every $T$~seconds. The prediction horizon is $N$ steps, i.e., $NT$~seconds. $\xi(t)$ denotes the continuous-time flat state at time $t$. $z_{i|k}$ denotes the predicted flat state at time $t=iT$, when the FTOCP is solved at time $t=kT$. We minimise a cost function over the sequence of flat states $\bm z_k = [z_{k|k}, z_{k+1|k}, ... , z_{k+N|k}]$ and flat control inputs $\bm v_k = [v_{k|k}, v_{k+1|k}, ..., v_{k+N-1|k}]$ subject to (A) the given dynamics, (B) initial and final constraints and (C) safety constraints.  The goal state is $\xi_g \in \R^{n+q}$, a flat state corresponding to a target state $x_g$ of the nonlinear system, i.e., $\xi(t) = \xi_g \implies x(t) = x_g$, where $(x(t), y(t)) = \Xi^{-1}(\xi(t))$. 

The FTOCP problem is the following optimization problem:
\vspace{-0.5cm}
\begin{subequations}
\begin{alignat}{3}
J^*(\xi(kT)) = ~&\underset{\bm{z}, \bm{v}}{\text{min.}} & \quad & \sum_{i=k}^{k+N-1}{l(z_{i|k}, v_{i|k} )} \label{eq:mpcCostFunction}\\
&\text{s.t.} &      & z_{i+1|k} = A_d z_{i|k} + B_d v_{i|k},   \label{eq:dyn_constraint}\\
& & & z_{k|k} - \xi(kT) \in \Dcal, \label{eq:init_cond_constraint} \\
& & & z_{k+N|k} = \xi_g, \label{eq:terminal_constraint}\\
& & & (z_{i|k}, v_{i|k}) \in \Hcal  \label{eq:state_constraint} \\
& & & \quad \forall i\in\{k, ..., k+N-1\}. \notag
\end{alignat}
\label{eqn:MPC_prob}
\end{subequations}
\textbf{\emph{Cost Function, \eqref{eq:mpcCostFunction}}}: $l(\xi, v)$ is the stage cost of action $v$ from a state $\xi$. We assume $l$ is convex in both arguments, is positive definite about $(\xi_g, 0)$, and is radially unbounded.

\textbf{\emph{Dynamics, \eqref{eq:dyn_constraint}}}: Under a zero-order hold, \eqref{eq:dyn_constraint} is the exact discretisation of the flat system \eqref{eqn:flat_system} i.e., $A_d = \exp{(A T)}, \quad B_d = \int_0^T \exp{(A \tau)} B d\tau$.

\textbf{\emph{Initial Condition, \eqref{eq:init_cond_constraint}:}} The initial state $z_{k|k}$, will be chosen by the FTOCP to be in the neighborhood of the flat state:
\eqn{
(z_{k|k} - \xi(kT)) \in \Dcal,}
where $\Dcal$ is an ellipsoid, and will be defined later, in~\eqref{eqn:Dcal}. Intuitively, $\Dcal$ represents the maximum tracking error between the reference state $\xi_{\textup{ref}}(t)$, and the current state $\xi(t)$.

\textbf{\emph{Final Condition, \eqref{eq:terminal_constraint}:}} The final state $z_{k+N|k}$ must be the goal state $\xi_g$, which is assumed to be a safe unforced equilibrium point for the system, i.e., $z = A_d z$.

\textbf{\emph{Safety, \eqref{eq:state_constraint}}}: Let $\Scal \subset \R^n$ be the set of safe states for the nonlinear system~\eqref{eqn:control-affine-system}. Then the set of safe flat states is ${\Ccal \subset \R^{(n+q)}}$, where $\Ccal = \{\xi \in \mathcal{N}: x \in \Scal, (x, y) = \Xi^{-1}(\xi) \in \mathcal{M} \}$.
Next, we define the set $\Hcal$ such that a flat state-input pair chosen in $\Hcal$ ensures that the intersample trajectory remains within a subset of $\Ccal$. Mathematically, ${\Hcal \subset \R^{n+q+m}}$ is s.t.:
\eqn{
(z_{i|k}, v_{i|k}) \in \Hcal \Rightarrow \xi_{\textup{ref}}(t) \in \Ccal \ominus \Dcal, \forall t\in [iT, (i+1)T)
\label{eqn:Hcal}}
where $\xi_{\textup{ref}}(t)$ is the solution to $\dot \xi_{\textup{ref}} = A \xi_{\textup{ref}} + B v_{i|k}$ from the initial condition $\xi_{\textup{ref}}(iT) = z_{i|k}$.  While computing $\Ccal$ and $\Hcal$ could be challenging, in many physical systems the safe set can be described in terms of the flat states. In these cases, constructing $\Ccal$ can be straightforward. To compute $\Hcal$, reachability-based methods for linear systems can be used, e.g.,~\cite{wu2020r3t}. The example below demonstrates an approach based on control barrier functions, e.g.,~\cite{breeden2021control, singletary2020control}. 

\textbf{\emph{The final result}}, by solving the FTOCP \eqref{eqn:MPC_prob}, is a sequence of flat states $\bm z_k$ and flat control inputs $\bm v_k$. The reference trajectory and input for the continuous-time flat system are defined for all $t \in [0, NT)$ using index $i \triangleq \operatorname{floor}(t/T)$ as
\begin{subequations}
\label{eqn:discrete_to_cont}
\begin{align}
\xi_{\textup{ref}}(t) &= e^{A (t-iT)} z_{i|k} + \left(\int_0^{t-iT} e^{A\tau} d\tau \right) B v_{i|k},\\
v_{\textup{ref}}(t) &= v_{i|k},\label{eqn:discrete_to_cont_control}
\end{align}
\end{subequations}

\begin{remark}
The FTOCP problem is a convex problem. Constraint \eqref{eq:init_cond_constraint} is a quadratic constraint, and the rest are linear constraints. As such, it is a second order cone program (SOCP), and can be solved efficiently using interior point methods or specialised solvers for MPC problems~\cite[Ch.~4,11]{boyd2004convex}\cite{ Alizadeh2003}.
\end{remark}

\begin{example}[Unicycle: High Level Planner\footnote{The code for the FTOCP is at \url{https://tinyurl.com/3c58cnjj}}]
Here, we provide details on computing $\Hcal$. Figure~\ref{fig:experiment_scenario} shows the safe set $\Scal$. Since the first two components of $\xi$ and $x$ are equal, $\Ccal = \{ \xi : [\xi_1, \xi_2] \in \Scal\}$. Since the tracking error lies in an ellipsoid $\Dcal$, we shrink the safe set by the diameter of $\Dcal$, and refer to this set as the \emph{tightened safe set}. This region is partitioned into 5 regions with overlapping boundaries, indicated by dashed rectangles. To find $\Hcal$, we constrain the trajectory due to a state-input pair $(z_{i|k}, v_{i|k})$ to lie within a single rectangular region for the next $T$ seconds. To do this, four sampled control barrier functions~\cite{breeden2021control} are defined, one for each edge of the rectangle: each edge defines a half space of the form $a^T \xi \leq b$. The barrier function is $h(\xi) = a^T \xi - b$, such that the safe set with respect to this edge is ${\{ \xi: h(\xi) \leq 0\}}$. A sufficient condition for $h(\xi(t)) \leq 0$ for all $t \in [(i+k)T, (i+k+1)T)$ is 
\eqnN{
h_0 \leq 0 \text{ and } h_0 + \dot h_0 T + 1/2 \max{\{0, \ddot h_0}\} T^2\leq 0,}
where $h_0 = a^T z_{i|k} -b$ and $\dot h_0, \ddot h_0$ are the first and second derivatives of $h$ at $z_{i|k}, v_{i|k}$. This is a linear constraint in both $z_{i|k}, v_{i|k}$, but requires integers to index the rectangles. Thus the FTOCP is a Mixed-Integer SOCP, which is solved online.

With regards to singularities, our method guarantees safety at all states where $\Xi$ is non-singular, (i.e., in the domain $\mathcal{M}$ of $\Xi$). For the unicycle, since the singular points correspond to states where the unicycle is at rest (see Example 1), these states are intrinsically safe. As such, in our planning problem, we can specify stopping at a desired location as a suitable target state. In general, independent analysis is needed to ensure the singular points are intrinsically safe. 
\end{example}

\subsection{Low-Level Tracking}

In this section, we construct a tracking controller. Consider a potentially time-varying, bounded matched disturbance $d \in \mathbb{L}^m_\infty$, ${\norm{d}_\infty \leq \bar d < \infty}$. The disturbed system is
\eqn{
\dot x = f(x) + g(x) u+ g(x) d(t).
\label{eqn:disturbed-control-affine-system}}
Transforming the disturbed system \eqref{eqn:disturbed-control-affine-system} to the flat space,
\eqn{
\dot \xi &= \frac{\partial \Xi}{\partial x} \dot x + \frac{\partial \Xi}{\partial y} \dot y 
= \frac{\partial \Xi}{\partial x} \big(f(x) + g(x) u + g(x) d(t)\big) + \frac{\partial \Xi}{\partial y} \dot y\notag\\
&= A \xi + B v + \frac{\partial \Xi}{\partial x} g(x) d(t),}
where $A,B$ are as in \eqref{eqn:flat_system}, and $\Xi$ is defined in \eqref{eqn:Xi}. Let $\xi_e \triangleq \xi - \xi_\textup{ref}$, $v_e \triangleq v - v_\textup{ref}$ and $w \triangleq \frac{\partial \Xi}{\partial x} g(x) d(t)$. Then
\eqn{
\dot \xi_e = A \xi_e + B v_e + w,
\label{eqn:flat_system_error_dynamics}
}
Assume the disturbance $w$ is bounded, $\norm{w}_\infty \leq \bar w$. Define the tracking feedback controller $\pi_e : \R^{n+q} \to \R^m$, and Lyapunov function
\eqn{
\pi_e(\xi_e) &= -\half R^{-1} B P \xi_e \label{eqn:riccati_controller}\\
V(\xi_e) &= \half \xi_e^T P \xi_e,
\label{eqn:lyapunov_eq}}
where $P \in \R^{(n+q)\times(n+q)}$ is a symmetric positive definite matrix satisfying a modified Riccati equation
\eqn{
PA + A^T P - P B R^{-1} B^T P = - Q - PP/\gamma^2,
\label{eqn:riccati}
}
where $Q \in \R^{(n+q)\times (n+q)}, R\in \R^{m\times m}$ are user-specified symmetric positive definite matrices. The parameter $\gamma > 0$ must be chosen such that $P$ exists. Since $(A, B)$ is controllable, $P$ exists for a sufficiently large $\gamma$~\cite{laub1979schur}.

\begin{lemma}
\label{lemma:V_is_ISSCLF}
The function $V$ \eqref{eqn:lyapunov_eq}, is an ISS-CLF for the flat error system \eqref{eqn:flat_system_error_dynamics}, wrt. the bounded disturbance $w$, $\norm{w}_\infty \leq \bar w$, under the feedback law~\eqref{eqn:riccati_controller}, with $\alpha$ and $\iota$ defined as
\eqn{
\alpha(\norm{\xi_e}) = \half \lambda_{min}(Q) \norm{\xi_e}^2, \quad \iota(\bar{w}) = \half \gamma^2 \bar{w}^2.
\label{eqn:alpha_iota}
}
\end{lemma}

\begin{proof}
The time derivative of $V$ along the closed-loop trajectories of system~\eqref{eqn:flat_system_error_dynamics} and feedback law~\eqref{eqn:riccati_controller}, is
\eqn{
\frac{d}{dt}V  &= \half \xi_e^T \left(-Q - \frac{1}{\gamma^2} PP \right)\xi_e + \xi_e^T P w\notag\\
&\leq- \half \xi_e^T Q \xi_e - \frac{\norm{P \xi_e}^2 }{2\gamma^2}  + \norm{P \xi_e} \norm{w}.
\label{eqn:vdot}
}
Adding and subtracting $\half \gamma^2 \norm{w}^2$ and factorizing yields
\eqn{\frac{d}{dt}V  &\leq - \half \xi_e^T Q \xi_e- \half \left(\frac{\norm{P \xi_e}}{\gamma}  - \gamma\norm{w}\right)^2 +\half \gamma^2 \norm{w}^2.}
Thus $V$ is a ISS-CLF \eqref{eqn:ISSCLF_def}, with $\alpha$, $\iota$ defined in~\eqref{eqn:alpha_iota}.
\end{proof}

\begin{lemma}
\label{lemma:control_invariance}
For the closed-loop flat error system \eqref{eqn:flat_system_error_dynamics}, under the feedback law~\eqref{eqn:riccati_controller},  bounded disturbances $w$, $\norm{w}_\infty \leq \bar w$, and $V$ as in~\eqref{eqn:lyapunov_eq}, the sub-level set 
\eqn{
\Dcal = \{ \xi_e : V(\xi_e) \leq V_{max}\},
\label{eqn:Dcal}}
is robustly control invariant, where
\eqn{
V_{max} = \half \gamma^2 \frac{\lambda_{max}(P)}{\lambda_{min}(Q)} \bar{w}^2.
\label{eq::V_max bound}}
\end{lemma}
\begin{proof}
Using $\dot V \leq -\alpha(\norm{\xi_e}) + \iota(\bar{w})$, we have $\alpha(\norm{\xi_e}) \geq \iota(\bar w) \implies \dot V \leq 0$, and therefore
\eqn{\norm{\xi_e}^2 \geq \gamma^2 \bar{w}^2 / \lambda_{min}(Q) \implies \dot V \leq 0.
\label{eqn:norm_xi}}
Similarly, since $V(\xi_e) \leq \frac{1}{2} \lambda_{max}(P) \norm{\xi_e}^2$, we have $V(\xi_e) \geq V_{max} \implies$~\eqref{eqn:norm_xi} and thus $\dot V \leq 0$, completing the proof.
\end{proof}

\begin{example}[Unicycle: Low Level Tracking]
\label{example:unicycle_low_level}
For the unicycle, we have $\bar w = \bar d$, since
\eqnN{
\bar w \triangleq \norm{\frac{\partial \Xi}{\partial x} g(x) d(t)}_\infty &= \norm{\bmat{1 & 0 & 0\\ 0 & 1 & 0 \\ 0 & 0 & 0 \\ 0 & 0 & 0} \bmat{\cos x_3 & 0 \\ \sin x_3 & 0 \\ 0 & 1} d(t)}_\infty\\
&= \norm{d(t)}_\infty \triangleq \bar d.}

The Riccati equation with $Q=I_4, R=I_2, \gamma=2$, defines $P$ in~\eqref{eqn:riccati}, the controller in~\eqref{eqn:riccati_controller} and $\Dcal$ in~\eqref{eqn:Dcal}, with ${V_{max} = 9.66 \bar w^2}$. 
\end{example}

\section{MAIN RESULT}
\label{section::main_result}

To summarise, the multirate controller is as follows. At the high-level, every $T$ seconds, the FTOCP \eqref{eqn:MPC_prob} is solved. Using~\eqref{eqn:discrete_to_cont}, the continuous-time reference trajectory $\xi_{\textup{ref}}(t), v_{\textup{ref}}(t)$ is determined. At the low-level, the flat error state $\xi_e$, and the flat input $v = v_{\textup{ref}} + \pi_e(\xi_e)$ are computed using~\eqref{eqn:riccati_controller}. Finally, the nonlinear control input is computed using~\eqref{eqn:dynfeedbackB}, $u = b(x, y, v)$, where $(x, y) = \Xi^{-1}(\xi)$.

We make the following assumptions and prove two lemmas before presenting the main result.
\begin{assumption}

The unmatched disturbances $w$ in the flat system~\eqref{eqn:flat_system_error_dynamics} are bounded, $\norm{w}_\infty \leq \bar w$.
\label{assum:bounded_flat_disturbance}
\end{assumption}
\begin{assumption}
\label{assum:full_info}
The state of the system (nonlinear and flat state) is perfectly measured at all times.
\end{assumption}
\begin{assumption}
\label{assum:initial_feasibility}
At the initial time $t_0$, \eqref{eqn:MPC_prob} is feasible.\footnote{Since FTOCP~\eqref{eqn:MPC_prob} is a convex program, the set of feasible  initial conditions can be computed explicitly~\cite{borrelli2017predictive}.}
\end{assumption}

\begin{remark} 
In the case of compact safe sets $\Scal$, as in many practical applications, Assumption~\ref{assum:bounded_flat_disturbance} can be verified, since $\partial \Xi/\partial x$ and $g(x)$ can be bounded, for instance using Lipschitz constants. See Example 3 for the closed form expression for $\bar w$ for the unicycle system.
\end{remark}
\begin{lemma}
\label{lemma:low_level_tracking}
Consider a reference trajectory ${(\xi_{\textup{ref}}(t), v_{\textup{ref}}(t))}$ defined over the interval $t\in[t_0, t_1]$ that satisfies the flat dynamics~\eqref{eqn:flat_system}. Under Assumption~\ref{assum:bounded_flat_disturbance}, if at $t_0$, the flat error $\xi(t_0) - \xi_{\textup{ref}}(t_0) \in \Dcal$, then the flat feedback control law
\eqn{
\pi(t, \xi) = v_{\textup{ref}}(t) + \pi_e(\xi(t) - \xi_{\textup{ref}} (t)),
\label{eqn:flat_control_input}
}
where $\pi_e$, \eqref{eqn:riccati_controller}, ensures tracking error remains within $\Dcal$:
\eqn{
\xi(t) - \xi_{\textup{ref}} (t) \in \Dcal, \quad \forall t \in [t_0, t_1].}
\end{lemma}
\begin{proof}
The proof is immediate since $\Dcal$ is a robust control invariant set for the flat error system (Lemma~\ref{lemma:control_invariance}).
\end{proof}

\begin{lemma}
Under Assumptions~\ref{assum:bounded_flat_disturbance}-\ref{assum:initial_feasibility},  and if the flat system is controlled using the tracking controller~\eqref{eqn:flat_control_input}, the FTOCP \eqref{eqn:MPC_prob}, is recursively feasible.
\label{lemma:recursive_feasibility}
\end{lemma}
\begin{proof}
Let the solution to~\eqref{eqn:MPC_prob} at the initial time $t_0$ be $\bm z_0, \bm v_0$. We show that
\begin{equation}\label{eq:feasSolution}
\begin{aligned}
\bm z_1 &= [ z_{1|0}, z_{2|0}, ..., z_{N-1|0}, z_{N|0}, \xi_g],\\
\bm v_1 &= [ v_{1|0}, v_{2|0}, ..., v_{N-1|0}, 0],
\end{aligned}
\end{equation}
is a feasible solution at the subsequent step $t=t_0 + T$. Since $z_{N|0} = \xi_g$ and $A_d \xi_g = \xi_g$, it is immediate that $\bm z_1, \bm v_1$ satisfies dynamics~\eqref{eq:dyn_constraint}, final condition~\eqref{eq:terminal_constraint} and safety requirement~\eqref{eq:state_constraint}. Constraint~\eqref{eq:init_cond_constraint} remains to be shown. From the solution $\bm z_0, \bm v_0$, the reference trajectory-input pair $\xi_{\textup{ref}}(t), v_{\textup{ref}}(t)$ for $t \in [t_0, t_0+NT]$ is computed~\eqref{eqn:discrete_to_cont}. Since $z_{0|0} = \xi_{\textup{ref}}(t_0)$, $\xi(t_0) - \xi_{\textup{ref}}(t_0) \in \Dcal$. By Lemma~\ref{lemma:low_level_tracking}, the low-level tracking controller guarantees that at the next step
\eqn{
\xi(t_0 + T) - z_{1|0} \in \Dcal,
}
since $\xi_{\textup{ref}}(t_0+T) = z_{1|0}$. Thus, the initial constraint~\eqref{eq:init_cond_constraint} is also satisfied, completing the proof.
\end{proof}

\begin{theorem}[Main Result]
\label{thm:main}
Consider the nonlinear system~\eqref{eqn:disturbed-control-affine-system}, subject to bounded, matched disturbances $d(t)$, $\norm{d}_\infty \leq \bar d$. Under Assumptions \ref{assum:bounded_flat_disturbance}- \ref{assum:initial_feasibility}, the proposed controller~\eqref{eqn:MPC_prob}, \eqref{eqn:discrete_to_cont}, \eqref{eqn:flat_control_input}, \eqref{eqn:dynfeedbackB} is  recursively feasible. Furthermore, the closed-loop system trajectories satisfy safety: $x(t) \in \Scal$ $\forall t \geq t_0$, and as $k\to \infty$, $x(kT) \in \{x: \exists y \text{ s.t. } \Xi(x, y) - \xi_g \in \Dcal\}$, i.e., a neighborhood of $x_g$ is reached.
\end{theorem}

\begin{proof}
\emph{First, we prove the recursive feasibility of the FTOCP}:
Since the low-level controller dictates $u$ such that the conditions of Lemma~\ref{lemma:recursive_feasibility} are met, recursive feasibility is maintained.

\emph{Second, we prove safety is maintained}:
By definition of $\Hcal$~\eqref{eqn:Hcal}, the FTOCP waypoints $\bm z$ and control inputs $\bm v$ are such that the reference trajectory $\xi_{\textup{ref}}(t)$ remains within the set $\Ccal \ominus \Dcal$. The low-level controller guarantees that the tracking error $\xi_e$ remains in $\Dcal$, i.e., $\xi(t) \in \Ccal$ for all $t\geq t_0$. Since $\xi \in \Ccal \implies x \in \Scal$, the nonlinear state remains safe. 

\emph{Finally, we prove convergence to goal state}:
Let $J^*(\xi(kT))$ be optimal cost at $t = kT$. Cost associated with the feasible solution~\eqref{eq:feasSolution} is $\bar J(\xi((k+1)T)) = J^*(\xi(kT)) - l(z_{0|0}, v_{0|0})$. Therefore, optimal cost at $t = (k+1)T$ satisfies
\begin{equation*}
\begin{aligned}
    J^*(\xi(kT)) & \geq \bar J(\xi((k+1)T)) = J^*(\xi(kT)
    ) - l(z_{0|0}, v_{0|0})\\
    & \geq J^*(\xi((k+1)T)).    
\end{aligned}
\end{equation*}
From the above equation and the positive definiteness of the stage cost $l$, the optimal cost $J^*(\xi(kT))$ is a Lyapunov function for the closed loop system~\eqref{eq:dyn_constraint},~\eqref{eqn:discrete_to_cont_control}. The result from Lemma~\ref{lemma:control_invariance} implies that $\lim_{k \rightarrow \infty} \xi(kT) \in \{\xi_g\} \oplus \Dcal$. Together with the invertibility of $\Xi$, this implies $x(t)$ reaches the neighborhood of $x_g$ defined in the theorem.
\end{proof}

\noindent \textbf{Discussion.}
The proposed controller differs from other planner-tracker controllers since the two levels are coupled through $\Dcal$, enabling formal guarantees on feasibility and safety. Furthermore, the controller is constructive given the diffeomorphism $\Xi$ and a few parameters.\footnote{The only parameters the user needs to specify are $N$, $T$, $l$, $Q$, $R$, and $\gamma$. A line search over $\gamma$ can be used to minimise the size of $\Dcal$.} The controller also provides greater flexibility in choosing $N, T$: any FTOCP feasible for some $(N, T)$ is also feasible for $(2N, T/2)$.


\section{Experimental Results}

\label{section::results}
We demonstrate the claims above using simulations and experiments on a ground rover and quadruped. The dynamical model and endogenous feedback of the unicycle model used, the high-level planner, and the low-level controller used in the experiment are described in Examples 1, 2, and 3 respectively.

\emph{Simulations.} The objective is to drive a unicycle from the start to the goal location~(Figure~\ref{fig:simulation_results}a,b). The system is subject to a matched disturbance, of magnitude $\sim10$\% of the control input magnitude. The reference trajectory always remains within the tightened safe set. While the unicycle's path can enter the margin between the safe set and the tightened set, it always remains safe. The value of the Lyapunov function~\eqref{eqn:lyapunov_eq}, remains below $V_{max}$ at all times, as expected~(Figure~\ref{fig:simulation_results}c).

\ifshowImages
\begin{figure}[t!]
    \centering
    \iffast
        \includegraphics[width=0.8\linewidth]{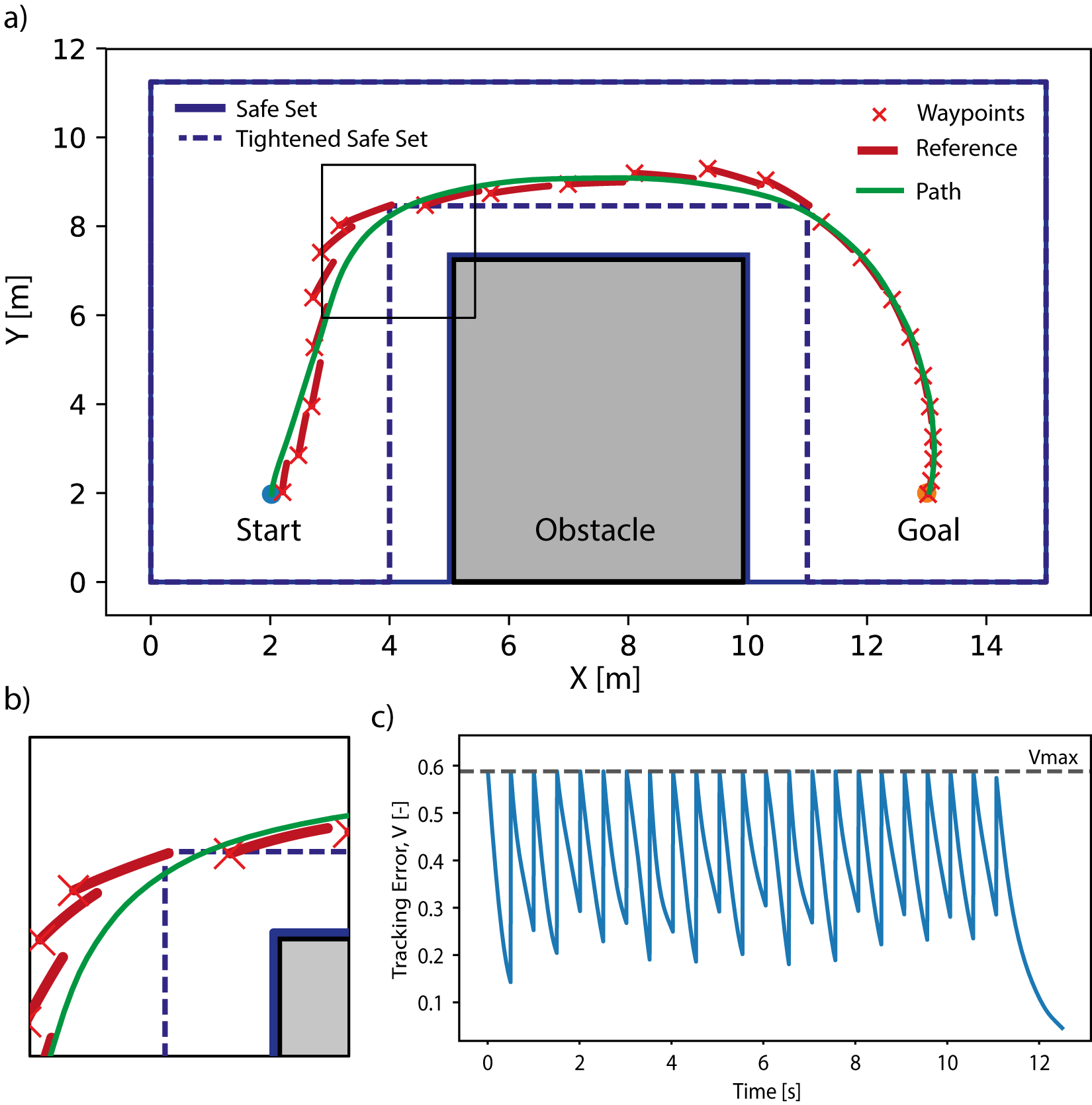}
    \else
        \includegraphics[width=0.8\linewidth]{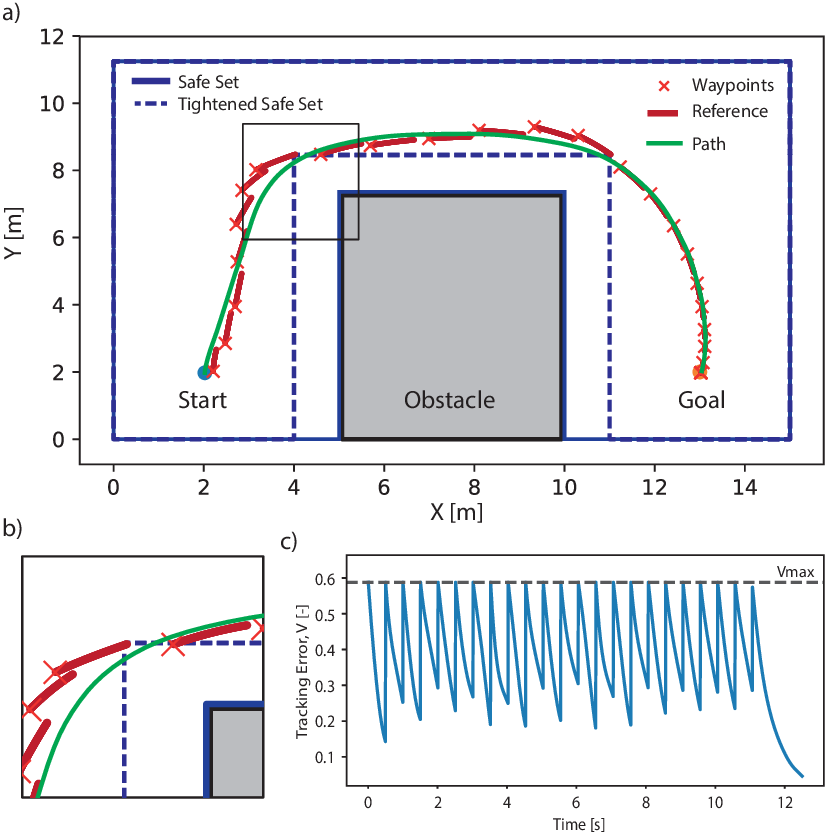}
    \fi
    \caption{Simulation results. (a) shows unicycle (green) and reference (red) trajectories. The reference is discontinuous, since it is recomputed every $T$~seconds. The start of each replanned reference trajectory is marked (red crosses). Black square is magnified in (b). (c) shows Lyapunov function against time, indicating that it remains below $V_{max}$.}
    \label{fig:simulation_results}
\end{figure}
\fi

\begin{figure}[t!]
    \centering
    \iffast
        \includegraphics[width=0.8\linewidth]{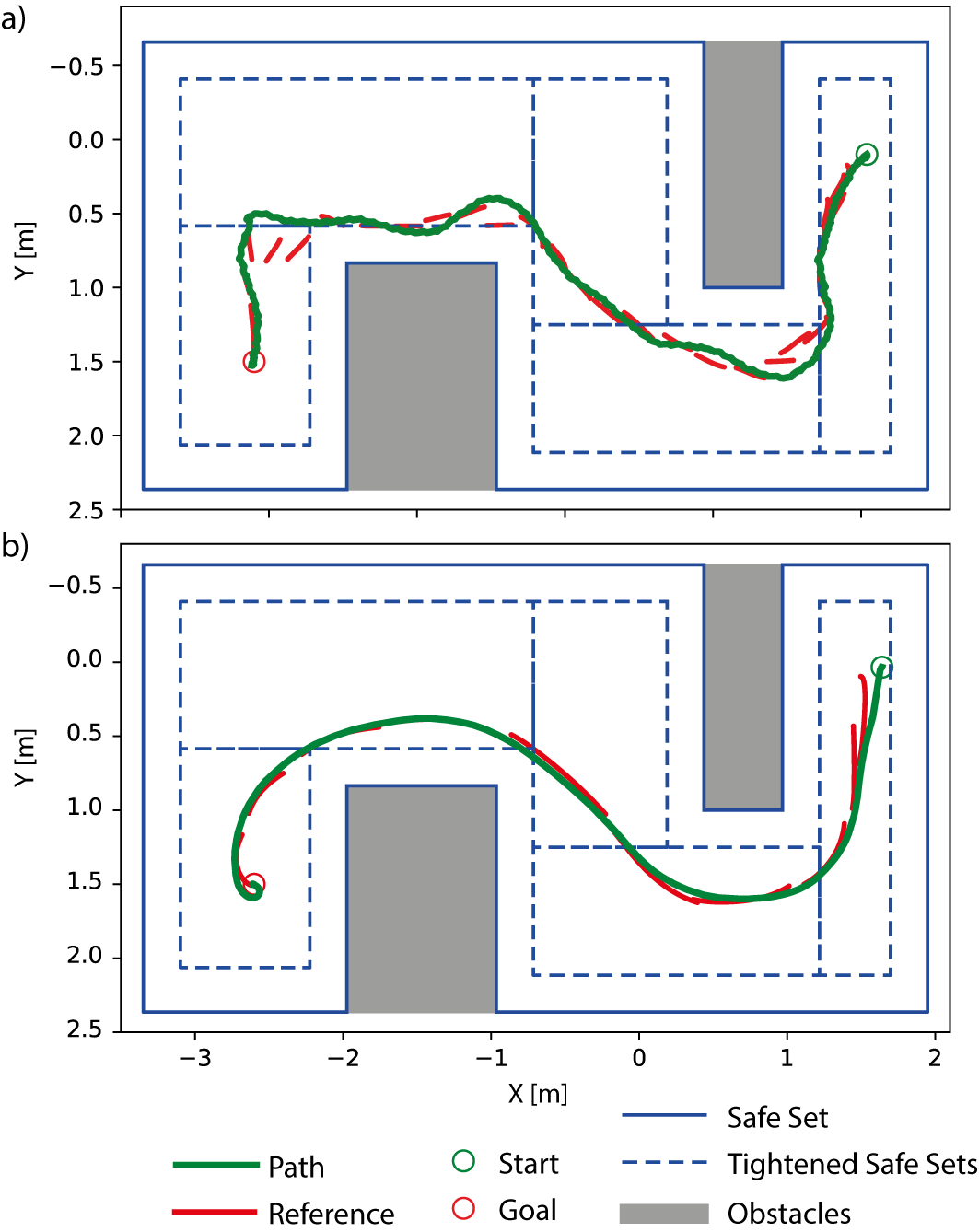}
    \else
        \includegraphics[width=0.8\linewidth]{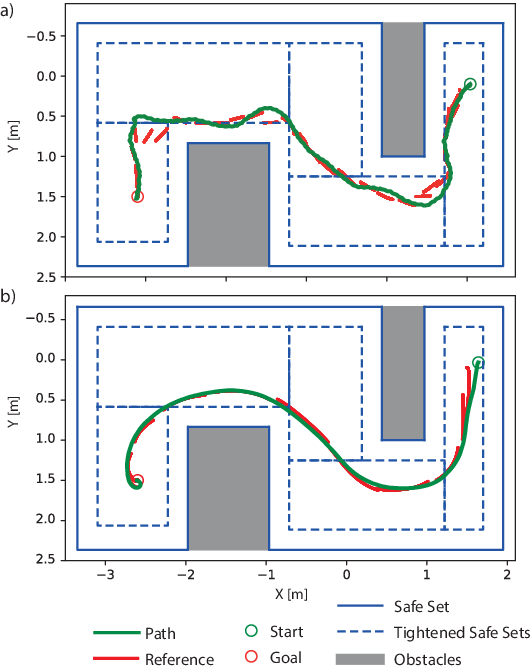}
    \fi
    \caption{Experimental Results. The quadruped (a) and the rover (b) navigate around gray obstacles in the environment to reach target location. See Figure~\ref{fig:experiment_scenario} for snapshots of the robots performing the experiments.}
    \label{fig:rover_map_plot}
\end{figure}

\emph{Experiments.} We show the real-time efficacy and robustness of our framework by implementing it on an AION R1 UGV rover, and a Unitree A1 quadruped (Figure~\ref{fig:rover_map_plot}). For both, the MI-QCQP MPC was implemented with \texttt{cvxpy} and solved using Gurobi. 
For the rover, we used $N = 9$, $T = 1.0$~seconds, and for the quadruped $N=30$, $T=2.0$~seconds. The low-level controllers were implemented digitally, running at 300 and 20~Hz for the rover and the quadruped respectively. Any error introduced by this sampling scheme is modeled as a part of $d$, the matched-disturbance to the dynamics.
Each iteration took between 0.05-0.2 seconds to replan. The communication and synchronization is done with ROS.  The voltage applied to the actuators are computed from the commanded velocities, by a PID for the rover, and an Inverse Dynamics Quadratic Program (ID-QP) designed using concepts in~\cite{buchli2009inverse} for the quadruped. Our method enables safe navigation despite the presence of modelling error arising due to inability of the robots' actuators to exactly track the commanded velocities.

\section{CONCLUSIONS}
\label{section::conclusion}
This paper details a constructive method to design a multirate controller for safety-critical differentially-flat systems. The coupling between the MPC and continuous controllers allows us to claim recursive feasibility of the MPC and safety of the nonlinear system.  Our theoretical claims were demonstrated in simulations and experiments. The effect of input constraints will be addressed in future works. We anticipate that penalising the flat inputs in the MPC cost function can improve input constraint satisfaction.

\section*{ACKNOWLEDGMENT}
The authors would like to thank Wyatt Ubellacker for his assistance with the quadruped simulation and experiments. 

\renewcommand{\baselinestretch}{0.81}

\bibliographystyle{IEEEtran}
\bibliography{references}

\end{document}